\documentclass[10pt,conference]{IEEEtran} 

\usepackage{amsmath,amsfonts}

\usepackage{wrapfig}
\usepackage{hyperref}
\usepackage{subcaption}
\newsavebox{\mybox}

\usepackage{amsthm}
\usepackage{mdframed}
\usepackage{centernot}
\usepackage{comment}

\newcommand\vd[2]{d_{i, p}}
\newcommand{\toolname}{\textsc{FairLay-ML}\xspace}
\usepackage{tikz}

\definecolor{gold}{rgb}{0.99,0.78,0.07}
\usetikzlibrary{arrows,shapes,snakes,automata,backgrounds,positioning,decorations.pathmorphing,
	decorations.markings,calc}

\tikzstyle{dtreenode}=[draw=blue!10!gray,rounded rectangle, minimum size=5mm,fill=blue!10!white]
\tikzstyle{dtreeleaf}=[draw=black!60,minimum width=1cm,minimum height=0.4cm,rectangle,fill=blue!50!white]
\tikzset{every loop/.style={looseness=7}}
\tikzset{
	gluon/.style={decorate,draw=black,
		decoration={coil,amplitude=1pt, segment length=5pt}}
}
\tikzset{
	gluon1/.style={decorate,draw=black,
		decoration={coil,amplitude=3pt, segment length=3pt}}
}
\tikzset{
	gluonew/.style={decorate,draw=black,
		decoration={coil,amplitude=1pt, segment length=2pt}}
}

\tikzset{bicolor/.style args={#1 and #2 and #3}{
		path picture={
			\tikzset{rounded corners=0}
			\fill [#1] (path picture bounding box.south west)
			rectangle
			($(path picture  bounding box.north west)!#3!(path picture bounding
			box.north east)$);
			\fill [#2]
			($(path picture bounding box.south west)!#3!(path picture bounding
			box.south east)$)
			rectangle (path picture bounding box.north east);
}}}

\tikzset{tricolor/.style args={#1 and #2 and #3 and #4 and #5}{
		path picture={
			\tikzset{rounded corners=0}
			\fill [#1] (path picture bounding box.south west)
			rectangle
			($(path picture  bounding box.north west)!#4!(path picture bounding
			box.north east)$);
			\fill [#2]
			($(path picture bounding box.south west)!#4!(path picture bounding
			box.south east)$)
			rectangle
			($(path picture  bounding box.north west)!#5!(path picture bounding
			box.north east)$);
			\fill [#3]
			($(path picture bounding box.south west)!#5!(path picture bounding
			box.south east)$)
			rectangle (path picture bounding box.north east);
}}}

\usepackage[many]{tcolorbox}
\tcbuselibrary{listings,skins}

\lstdefinestyle{mystyle}{
  xleftmargin=0pt,
   basicstyle={\footnotesize\ttfamily},
   aboveskip=3mm,
   belowskip=3mm,
   keywordstyle=\bfseries,
   showstringspaces=false,
  escapechar=?,
  language=Java
}
\definecolor{code_indent}{HTML}{CCCCCC}

\newtcblisting{mylisting}[2][]{
  arc=0pt, outer arc=0pt,
  listing only,
  listing style=mystyle,
  title={\large #2},
  #1
}

 \definecolor{dkgreen}{rgb}{0,0.6,0}
 \definecolor{gray}{rgb}{0.5,0.5,0.5}
 \definecolor{mauve}{rgb}{0.58,0,0.82}

\definecolor{cadmiumgreen}{rgb}{0.0, 0.42, 0.24}
\definecolor{verde}{rgb}{0.25,0.5,0.35}
\definecolor{jpurple}{rgb}{0.5,0,0.35}
\definecolor{darkgreen}{rgb}{0.0, 0.2, 0.13}

\usepackage[ruled,vlined,linesnumbered,lined]{algorithm2e}
\usepackage{algpseudocode}

\usepackage{mathtools,xparse}

\usepackage{tabu}
\usepackage{multirow}

\usepackage{pifont}
\usepackage{enumerate}
\usepackage[shortlabels]{enumitem}

\usepackage{pgfplotstable}
\usepackage{pgfplots}

\usepackage[noframe]{showframe}
\usepackage{framed}

 {\endMakeFramed}
 \definecolor{shadecolor}{gray}{0.85}

\definecolor{bgblue}{RGB}{245,243,253}
\definecolor{ttblue}{RGB}{91,194,224}

\mdfdefinestyle{mystyle}{%
  rightline=true,
  innerleftmargin=10,
  innerrightmargin=10,
  outerlinewidth=3pt,
  topline=false,
  rightline=false,
  bottomline=false,
  skipabove=\topsep,
  skipbelow=false
}

\newtcolorbox{myboxi}[1][]{
  breakable,
  title=#1,
  colback=white,
  colbacktitle=white,
  coltitle=black,
  fonttitle=\bfseries,
  bottomrule=0pt,
  toprule=0pt,
  leftrule=3pt,
  rightrule=3pt,
  titlerule=0pt,
  arc=0pt,
  outer arc=0pt,
  colframe=black!50,
}

\newtcolorbox{myboxii}[1][style=mystyle]{
  breakable,
  freelance,
  colback=white,
  colbacktitle=white,
  coltitle=black,
  fonttitle=\bfseries,
  bottomrule=0pt,
  boxrule=0pt,
  colframe=white,
  after skip=0pt,
  overlay unbroken and first={
    \draw[white!75!black,line width=3pt]
    ([yshift=-9pt]frame.north west) --
    ([yshift=9pt]frame.south west);
  },
}
\usepackage{graphicx} 

\title{FairLay-ML: Intuitive Debugging of Fairness in Data-Driven Social-Critical Software}

\author{
\IEEEauthorblockN{\textit{Normen Yu}}
\IEEEauthorblockA{
\textit{Dept. of Computer Science}\\
\textit{Pennsylvania State University}\\
\textit{normenyu@gmail.com} \\
}
\and
\IEEEauthorblockN{\textit{Luciana Carreon}}
\IEEEauthorblockA{
\textit{Dept. of Computer Science}\\
\textit{University of Texas at El Paso}\\
\textit{lkcarreon2@miners.utep.edu} \\
}
\and
\IEEEauthorblockN{\textit{Gang Tan}}
\IEEEauthorblockA{
\textit{Dept. of Computer Science}\\
\textit{Pennsylvania State University}\\
\textit{gtan@psu.edu} \\
}
\and
\IEEEauthorblockN{\textit{Saeid Tizpaz-Niari}}
\IEEEauthorblockA{
\textit{Dept. of Computer Science}\\
\textit{University of Texas at El Paso}\\
\textit{saeid@utep.edu} 
}
}
\begin{document}

\maketitle

\begin{abstract}
Data-driven software solutions have significantly been used in critical domains with significant socio-economic, legal, and ethical implications. The rapid adoptions of data-driven solutions, however, pose major threats to the trustworthiness of automated decision-support software. A diminished understanding of the solution by the developer and historical/current biases in the data sets are primary challenges. To aid data-driven software developers and end-users, we present \toolname, a debugging tool to test and 
explain the fairness implications of data-driven solutions. \toolname visualizes the logic of datasets, trained models, and decisions for a given data point. In addition, it trains various models with varying fairness-accuracy trade-offs. Crucially, \toolname incorporates counterfactual fairness testing that finds bugs beyond the development datasets. We conducted two studies through \toolname that allowed us to measure false positives/negatives in prevalent counterfactual testing and understand the human perception of counterfactual test cases in a class survey. \toolname and its benchmarks are publicly available at~\url{https://github.com/Pennswood/FairLay-ML}. The live version of the tool is available at~\url{https://fairlayml-v2.streamlit.app/}. We provide a video demo of the tool at~\url{https://youtu.be/wNI9UWkywVU?t=133}. 

\end{abstract}

\section{Introduction}
\label{sec:intro}
Addressing systemic bias and unfairness has been a fundamental societal problem, and nothing is more ``systemic" in 21st century than life-altering decisions made by automated algorithms that learn their logic from historical data.
These automated decision-making algorithms act as gatekeepers for many social decisions: should a person remain in prison~\cite{compas-software}, is a person qualified for a job~\cite{Amazon}, and will a loan receiver become a default~\cite{LoanDefault}.
This is concerning as many of these algorithms are becoming less human-driven and human-understandable~\cite{lecun2015deep}. The problem of human understandability and explainability exacerbates discrimination in data-driven automated algorithms since some stakeholders (e.g., software developers, domain experts, end users, etc.) can find it more difficult to debug unfairness---the process of detecting, localizing, and fixing fairness bugs. This paper presents \toolname, an intuitive debugging tool that aids data-driven software developers and end users to understand fairness implications of their solutions.  

There are plenty of software tools in the literature to improve fairness~\cite{bellamy2019ai,bird2020fairlearn}. IBM AIF360~\cite{bellamy2019ai} and Fairlearn~\cite{bird2020fairlearn} provide metrics and algorithms to measure unfairness and mitigate it via pre-processing (e.g., modifying training datasets), in-processing (e.g., modifying the loss objective), and post-processing (e.g., calibration of scores) mechanisms. 
Google What-if toolkit~\cite{8807255} is the closest to \toolname. Both tools focus on individual fairness rather than group fairness, that is to ensure similar individuals receive similar outcomes. Google What-if toolkit helps find a discriminatory instance by comparing a base data point to its similar counterfactual with a different protected attribute. \toolname, instead, goes beyond the training and development set and adapts individual fairness testing methods~\cite{galhotra2017fairness,ICSE2022} to explore the arbitrary input space to find discriminatory instances. \toolname also allows users to modify test cases and define counterfactuals based on their intuitions (e.g., characteristics that make two individuals from different groups similar).  

We posit that significant user interactions in training and validating data-driven solutions against fairness requirements are critical design goals. Hence, \toolname integrates automation with human feedback, e.g., we adapt automated test-case generation, allowing users to issue manual test cases. Throughout all activities, we provide explanations and visualizations guiding users to make informed decisions.  

\toolname is a GUI app that integrates multiple research ideas into a unified toolkit.
\toolname takes a dataset of a target application (income, loan, recidivism, etc.)
as input and (i) visualizes interactions between different features (e.g., distribution of job titles by gender) via plotly and streamlit, (ii) runs a multi-objective evolutionary algorithm based on \textsc{Parfait-ML}~\cite{ICSE2022} to train multiple models of varying accuracy vs. fairness based on a given threshold, (iii) explains the model logic based on the importance of different features on the outcome, (iv) generates base test cases that may lead to individual discrimination, (v) explains the model's decision for a given test case, and (vi) allows users to create and modify counterfactuals of a given base test case.

\section{Background}
\label{sec:background}
Fairness is an emergent concern as Machine Learning (ML) software is rapidly evolving, deployed to assist in high-stakes decisions such as in hiring~\cite{Amazon}, parole~\cite{compas-article}, and financial lending~\cite{Home-Credit-Default-Risk}.
ML models are often trained with large datasets containing historical or systematic biases, and can lead models to maintain or amplify existing disparities, generating unfair outcomes based on legally protected attributes (e.g., gender, race). For example, the recidivism risk assessment tool, COMPAS, has yielded biased predictions that disadvantage African Americans~\cite{compas-article} and Amazon had to deactivate a hiring algorithm that was biased against female applicants~\cite{Amazon}.

\noindent\textbf{Notion of Individual Fairness.} Fairness through awareness ($FTA$)~\cite{dwork2012fairness} requires two \textit{individuals} deemed to be similar (based on their non-protected attributes) to be treated similarly (regardless of their protected attributes). However, defining the notion of similarity between individuals is difficult. Counterfactual fairness~\cite{kusner2017counterfactual} uses a causal dependency diagram between variables and requires that the ML outcome is causally independent of protected attributes.
Here, we focus on a simpler notion, \textit{counterfactual discrimination} as proposed by Galhotra et al.~\cite{galhotra2017fairness} that requires the ML outcome to remain similar between an individual and their counterfactual who has only one different protected value. This notion is the primary oracle of fairness testing~\cite{galhotra2017fairness}.

\section{\toolname}
\label{sec:tool-description}

\begin{figure}
    \centering
    \includegraphics[width=0.48\textwidth]{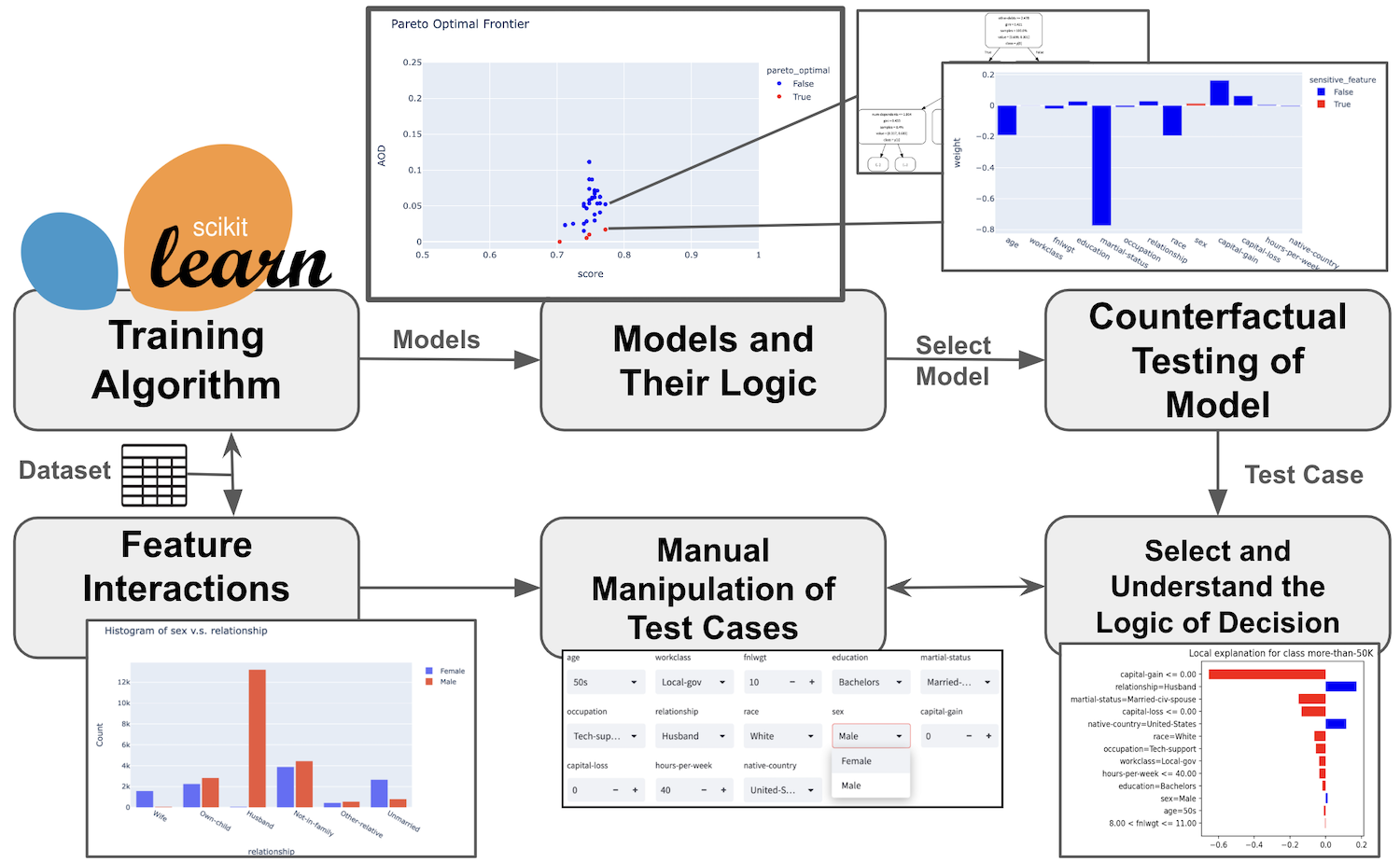}
    \caption{\toolname Usage.}
    \label{fig:tool-usage}
    \vspace{-1.0 em}
\end{figure}

We present \toolname, a GUI MLOps pipeline that allows users to upload data and/or machine learning models to our toolkit for fairness debugging.
\toolname identifies, explains, and suggests remedies to improve the model fairness. \toolname implements the remedies, retrains the model on the given dataset, and allows the user to re-evaluate remedies for accuracy and fairness, and save the upgraded dataset and model, all within our pipeline. 
One key component of \toolname is counterfactual testing that measures the unfairness of data-driven software by computing the differences between an original test case and its counterfactual with different protected attributes. 
In doing so, \toolname provides various data visualizations on interactions between different attributes, relationships between configurations of training algorithms and fairness, and the logic of ML model in inferring a decision.
This enables users to generate test cases (e.g., base and counterfactual test cases) that are natural and reflect their perceptions of real-world relationships.   

Figure~\ref{fig:tool-usage} shows the end-to-end usage of \toolname. A user begins by uploading their tabular data (both in the original form with labeled categories and features, as well as its numerically encoded) and a training algorithm (e.g., decision trees, logistic regressions, etc.). The user specifies the sensitive attribute and categories within the sensitive attribute they wish to study. From the dataset, we first create an 80/20 train-test-split to have training and validation datasets. \toolname also provides histograms and correlation metrics between each column and the sensitive attribute on the training data (feature interactions). \toolname uses \textsc{Parfait-ML}~\cite{ICSE2022} to generate various models with varying fairness/accuracy trade-offs. The visualization allows users to investigate different models where \toolname plots the logic of models (e.g., normalized weights of contributions from different features).
\toolname highlights Pareto-optimal models with red dots---those that achieve an ideal accuracy vs. fairness trade-off. Once the user picks one model as the decision-making model, \toolname performs counterfactual testing based on the \textit{counterfactual discrimination} notion using \textsc{Themis}~\cite{galhotra2017fairness}. The tool plots the generated test cases and highlights four types of data points:
i) the original and counterfactual data points agree on a positive outcome; ii) the original and counterfactual data points agree on a negative outcome; iii) the original data point classified as positive, while the counterfactual one is classified as negative; and iv) the original data point classified as negative, while the counterfactual one classified as positive. The user can then investigate different data points. Once they click on one data point, \toolname shows the local explanations on how the model generates a particular decision using \textsc{LIME}~\cite{lime}. \toolname presents a dashboard to users who can manipulate the data point to see how different changes contribute to the outcome, enabling users to investigate counterfactual test cases. They can use the LIME plot and identify which features/values might lead to a realistic counterfactual test case with significant discrimination.
Next, we overview the key components of \toolname.

\vspace{0.1 em}
\noindent \textbf{Feature Interactions.}
The data visualization tools provide an overview of how features correlate to the sensitive feature, helping users explore the distribution of classifications for each feature, and understand how the features in a dataset are interconnected in the context of ML bias. \toolname’s data visualization tool identifies key features to pay attention to when analyzing bias. For example, zip code has often been found to accidentally be used by ML algorithms as a proxy for race, as can more obvious bugs in the pipeline such as ``husband" and ``wife" being used as a proxy for gender. 

Our tool provides a user interface to dive deeper into correlations for each feature. By clicking on the correlation graph directly, users are shown a histogram of how likely each possible value of a specific feature to appear in a protected class of a sensitive feature. For example, Figure~\ref{fig:tool-usage} includes `Feature Interactions' where the attribute sex is ``female'', the histogram of ``husband" relationship has no density. 
This gives users better context on whether specific values of a feature could be used when they modify a test case. 

\vspace{0.25 em}
\noindent \textbf{Model Visualization and Explanation.}
\toolname displays models' logic to gain (global) insights into how features are used to make decisions. \toolname supports displaying decision tree and random forest models by plotting all tree structure directly, while it shows the weights of each feature or the slope of the hyperplane with respect to each feature for other algorithms. Finally, \toolname explains how a model classifies specific data points using local interpretable model-agnostic explanations called LIME~\cite{lime}. LIME shows how each feature value of a data point impacts the model's decision, enabling users to understand what features and values may change outcomes (e.g., from favorable unfavorable). These methods of visualization also combine with feature interactions to provide explanations of the form ``This model places a high weight on feature category X, and X is strongly correlated with the sensitive feature, causing unfairness." The `Select and Understand Model Logic' box in Figure~\ref{fig:tool-usage} shows how different features and their values contribute to the outcome of decision. 

\vspace{0.25 em}
\noindent \textbf{Counterfactual Test Case Generation.}
Given a model and a set of test data points for the model, \toolname can scatterplot data points for their predicted probability and highlight counterfactual points to help during testing to see which specific individuals are impacted by unfairness. For example, some models may only generate counterfactuals on the decision boundary (when the predicted probability is close to 0.5). Some models may have a bigger spread of counterfactuals, indicating wider systemic bias.

Users can explore individual data points and filter specific counterfactuals even if it generated a more or less accurate prediction. Users can filter by false positive, false negative, true positive, or true negative. \toolname then uses previously mentioned explanation techniques (e.g. LIME and model visualization) to generate a common story for the user. The `Manual Manipulation of Test Case' box in Figure~\ref{fig:tool-usage} shows an example of interactive counterfactual test case generation.

\section{EMPIRICAL EVALUATION}
\label{sec:results}

This preliminary evaluation of \toolname for debugging counterfactual test cases asks these research questions (RQs):

\begin{itemize}
    \item \textbf{RQ1.} What is the false positive and false negative rates of prevalent counterfactual testing? 
    \item \textbf{RQ2.} How accurately do humans create counterfactuals?  
\end{itemize}

Currently, \toolname supports four datasets~\cite{UCI-DS}: Adult Census Income, Credit Bank, Loan Default, and COMPAS over three protected attributes (sex, race, age). Since Adult Census Income is the most intuitive and simple applications (i.e., predicting whether incomes are above 
\$50k based on personal characteristics like the job, hours-per-week, sex, race, etc.), we will focus on this dataset to answer RQs. 

\begin{figure}
    \centering
    \includegraphics[width=0.35\textwidth]{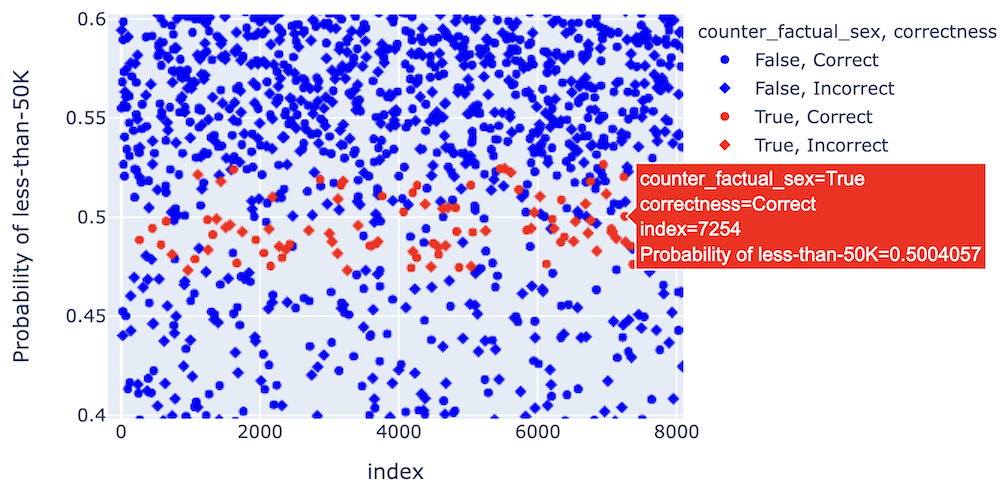}
    \caption{Visualization of Counterfactual Test Case Generations. }
    \label{fig:counterfactual-study}
    \vspace{-2.0 em}
\end{figure}

\vspace{0.5 em}
\textit{RQ1. False Positives/Negatives of Counterfactual Testing.}
\toolname automatically generates test cases via a causal method, similar to  \textsc{Themis}~\cite{galhotra2017fairness}.
The approach is to find \textit{individual discrimination} (ID) instances---two inputs that share the same relevant inputs, but they receive two different outcomes due to their differences in the protected attributes---in a given amounts of time.
Figure~\ref{fig:counterfactual-study} shows 8,000 test cases visualized by \toolname. The application is the Adult Census Income and \textit{sex} is the protected attribute.
The red points show ID instances where diamonds and square show whether the original instance or counterfactual one received a favorable outcome (an income above \$50k).
However, simply perturbing protected attributes from $A$ to $B$ can lead to false positives or negatives. For example, an ID instance generated by changing the protected attribute from Male to Female may be false without adjusting the relationships (e.g., Husband to Wife).
Similarly, a benign test case may be buggy once all the relevant features are adjusted for it.
We study the false positive and negative rates of prevalent counterfactual testing with only perturbing protected attributes. The key feature of \toolname is that it allows to pick a test case and adjust any relevant features to generate a valid and proximal counterfactual. 
Thus, we use \toolname to evaluate false positive and false negative rates. We pick 100 test cases uniformly at random from the set of test cases, and then manually evaluate their naturalness.
We observe that the rate of false positives, false negatives, true positives, and true negatives are 20.4\%, 27.5\%, 72.5\%, and 79.6\%. An example of false positive is  \texttt{<sex="Male", race="Black", occupation="Cleaner", relationship="Husband", hours\_week=80, country="Unknown">} where changing the sex attribute to ``Female" flips the outcomes from favorable (more than \$50k) to unfavorable (less than \$50k); but once the relationship is adjusted for (``Husband" $\to$ "Wife"), then both original and counterfactual outcomes become favorable. 

\begin{figure*}
    \centering
    \includegraphics[width=0.75\textwidth]{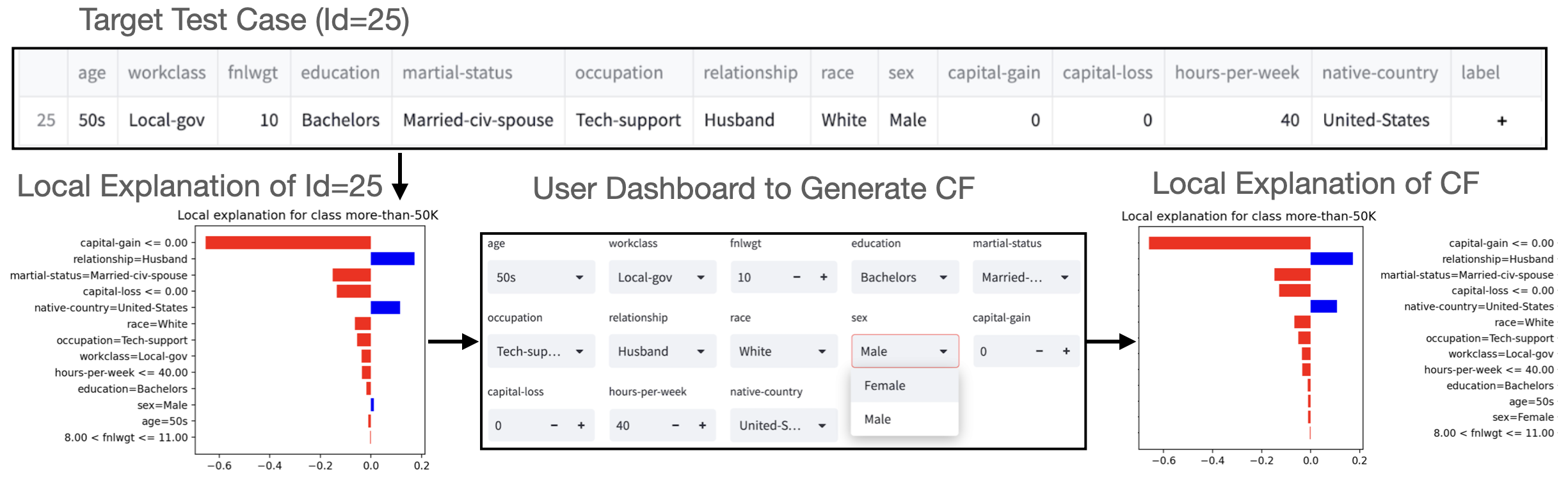}
    \caption{Steps in Generating Counterfactuals (CF) with Human Intuition}
    \label{fig:survey-CF}
    \vspace{-1.0 em}
\end{figure*}

\vspace{0.5 em}
\textit{RQ2. Validity and Proximity of Human Generated Counterfactuals.}
\toolname allows us to perform a study to understand the perception of counterfactuals
by humans. 
We perform a class survey with $17$ participants (UG and graduate students from the UTEP)
where they use \toolname to modify a sample (e.g., the occupations from a private company to a non-profit organization) to generate a counterfactual with a different background based on their perception of validity (a natural counterfactual that adheres to real-world constraints) and proximity (a minimal number of changes to the original sample). 
Figure~\ref{fig:survey-CF} shows the steps to complete the survey where
students can change up to 7 features of the target test cases (after changing
the sex attribute from male to female) and report
a combination that flips the ML decision from a favorable outcome (for the male individual)
to an unfavorable one (for the female individual) by querying the model using \toolname.
The survey asked each student to evaluate the CFs reported by one or two other students
in terms of its proximity to the target test case and validity w.r.t real-world constraints. The results showed that 9 students out of 17 were able to find a combination that changes the ML outcome with a different sex and relationship. The number of changes vary between 2 and 7. The nine valid responses received 12 feedback on the validity and proximity. All the responses said "yes" to the validity of CFs, implying that the counterfactual adheres to the real-world constraints. However, 6 of those evaluation feedback said that the CFs are not proximal to the original target test case (e.g., a counterfactual changed the education level). Most students tried to change race to black and believed that such changes may have significant impacts on the outcome, while being valid/proximal. We also observed that counterfactuals vary between participants significantly. For example, some participants said a separated individual may still have wife/husband relationship. Others believed that changing the country of origin does not preserve the proximity of CFs. 
\toolname allowed i) students to easily make changes to a sample and query the model with arbitrary samples; and ii) authors to visualize the interactions between variables to evaluate validity/proximity of counterfactuals.

\section{Limitations}
\label{sec:limits}

Our solution, \toolname, currently supports tabular data in the form of a CSV and binary classification algorithms from scikit-learn such as Decision Trees, SVMs, Random Forest, and Logistic Regression.

\section{Related Work}
\label{sec:related}
\textsc{Seldonian}~\cite{Seldonian} presented a mitigation technique that allows users to directly specify (arbitrary) undesirable behaviors.
The key innovation of \textsc{Seldonian} is to enforce safety requirements during training in the offline contextual bandits setting with probabilistic guarantees.
\toolname, on the other hand, focuses on helping users detect and understand discrimination.  
While eXplainable AI (e.g., LIME~\cite{lime}, SHAP~\cite{lundberg2017unified}, etc.) explain how ML models globally (for all data points) and locally (for a specific data point) make decisions, \toolname focuses specifically on fairness and allows users to identify significant test cases and compare data points. 

\section{Future Work}
\label{sec:conclusion}
We presented \toolname, an intuitive debugging tool to understand the implications of the fairness of sociotechnological software. 
One exciting future direction is to support mitigating unfairness. 
One idea that is partially implemented in the current tool is to mask features which are highly correlating with protected attributes. 

\vspace{0.5 em}
\noindent \textbf{Acknowledgement.} Carreon was supported by NSF grant CNS-2137791 through CAHSI REU program. This research has been supported by NSF under Grant No. CNS-2230060 and CNS-2230061.

\bibliography{references}
\bibliographystyle{IEEEtran}

\end{document}